\documentclass[runningheads]{llncs}
\usepackage[T1]{fontenc}
\usepackage{graphicx,verbatim}
\usepackage{amsfonts}
\usepackage{amsmath}
\usepackage{booktabs}
\usepackage{xcolor}
\usepackage[misc,geometry]{ifsym}
\begin{document}
\title{LoRCA: LoRA Cycle Adaptation for Histology to HiP-CT Translation with DINOv3}
{}

\titlerunning{LoRCA}
%
\author{Yang Zhou\inst{1}\textsuperscript{*} \and
Edoardo Occhipinti\inst{1}\textsuperscript{*} \and
Banboye Kidzeru Elvis\inst{2,3,4} \and
Jishizhan Chen\inst{1} \and
Stathis Megas\inst{5,6,7,8} \and
Joseph Brunet\inst{1,9} \and
Joanna Purzycka\inst{9} \and
Theresa Urban\inst{9}\and
Hector Dejea\inst{9} \and
Sarah Amalia Teichmann\inst{5,10,11} \and
Menna R Clatworthy\inst{2,3,4,12} \and
Paul Tafforeau\inst{9} \and
Peter D Lee\inst{1} \and
Claire L Walsh\inst{1}\textsuperscript{\Letter}
}
\authorrunning{Y. Zhou et al.}
%
\institute{Multiscale X-ray Imaging Lab, Department of Mechanical Engineering, University College London, UK \\
    \email{\{yang.zhou,e.occhipinti,jishizhan.chen,j.brunet,
    peter.lee,c.walsh.11\}@ucl.ac.uk} \and
    Molecular Immunity Unit, Department of Medicine, University of Cambridge, UK \and
    Cambridge Institute of Therapeutic Immunology and Infectious Diseases, University of Cambridge, UK \and
    Cellular Genomics, Wellcome Trust Sanger Institute, UK \\
    \email{be3@sanger.ac.uk}\and
    Cambridge Stem Cell Institute, Jeffrey Cheah Biomedical Centre, Cambridge Biomedical Campus, University of Cambridge, Cambridge, UK \and
    Centre for AI in Medicine, Department of Applied Mathematics and Theoretical Physics, University of Cambridge, UK \and
    Cavendish Laboratory, Department of Physics, University of Cambridge, UK \and
    Medical University of Vienna, Austria \\
    \email{efstathios.megas@meduniwien.ac.at} \and
    European Synchrotron Radiation Facility, France \\
    \email{\{joanna.purzycka,theresa.urban,paul.tafforeau\}@esrf.fr, hector.dejea@anaxam.ch}\and
    Department of Medicine, University of Cambridge, Cambridge, UK \and
    CIFAR Macmillan Multi-scale Human Programme, CIFAR, Toronto, Canada \\
    \email{sat1003@cam.ac.uk}\and
    Cambridge University Hospitals NHS Foundation Trust and NIHR Cambridge Biomedical Research Centre, UK \\
    \email{mrc38@medschl.cam.ac.uk}
}

\maketitle              

\begingroup \renewcommand\thefootnote{} \footnotetext{* Equal contribution} \footnotetext{\Letter Corresponding author: c.walsh.11@ucl.ac.uk} \footnotetext{Paper is accepted in MICCAI 2026 SASHIMI Workshop} \endgroup

\begin{abstract}

Hierarchical Phase-Contrast Tomography (HiP-CT) is a synchrotron-based X-ray imaging technique that enables non-destructive, volumetric imaging of intact organs with multi-resolutions bridging 20 $\mu m$/voxel for whole organs to near-cellular resolution ($\sim$0.8 $\mu m$/voxel) in local regions. This offers the opportunity to bring volumetric whole-organ context to histology. However, nonlinear registration between H\&E histological sections and HiP-CT volumes is challenging due to the substantial differences in feature representations. Direct alignment is hindered by the appearance mismatch between grayscale HiP-CT and RGB histology. Synthesis-before-registration methods have shown strong results in histology-to-MRI and histology-to-CT alignment. However, existing approaches either rely on manual anatomical contours or are trained from scratch without semantic constraints, limiting their generalisability to soft tissue organs and novel modalities. We propose LoRCA (LoRA Cycle Adaptation), a cycle-consistent style translation framework built on a shared frozen DINOv3 backbone with modality-specific LoRA adapters, learning modality-specific representations that are decoded and adversarially trained. LoRCA enables structure-preserving translation without requiring paired training data. The frozen backbone is intended to act as a structural anchor that prevents content drift by preserving pretrained semantic-extraction capability. We evaluate translation quality using Fréchet Inception Distance (FID) and structural fidelity via mutual information and Canny edge preservation. LoRCA outperforms CycleGAN in both translation quality and structural consistency. As a preliminary indicator of downstream registration utility, we find that style-translated images yield increased feature correspondences under MatchAnything on manually aligned HiP-CT and histology test pairs, suggesting that LoRCA-style translation is a promising step towards the registration of 2D histological sections to 3D HiP-CT volumes.


\end{abstract}
\section{Introduction}
Understanding the three-dimensional cellular structure of intact human organs is a key challenge in biomedical research. Histology remains the gold standard for cellular-level tissue analysis \cite{bai2023deep}, as it offers rich morphological details through various staining protocols. However, histology preparation is inherently destructive and produces isolated two-dimensional slices, making it difficult to correlate with 3D anatomical context \cite{chen20252d}. This restricts the studies of spatially distributed structures such as renal glomeruli and vasculature. Hierarchical phase-contrast tomography (HiP-CT) has recently emerged as a non-destructive imaging modality \cite{walsh2021imaging}, enabling multi-scale volumetric imaging of intact human organs at 20um/voxel for whole organ overviews to near cellular resolution (ca. 0.8 $\mu m$/voxel) in local regions. The multiple resolution 3D scans are intrinsically aligned to one another. HiP-CT has demonstrated the ability to resolve individual glomerulus segmentation across the whole human kidney \cite{zhou2026multiscale}, offering an insight into their 3D organisation. Despite these advances, fusion of HiP-CT with histology remains challenging, as no standard workflows and registration pipelines exist for correlating HiP-CT volumes with downstream histology.

A natural approach to bridging this gap is multi-modal registration. However, this is challenging due to the strong appearance mismatch between 3D grayscale HiP-CT and 2D RGB histology, as well as distortions introduced during tissue sectioning. A common strategy is the synthesis-before-registration paradigm \cite{iglesias2018joint}, which has been explored in histology-to-MRI alignment \cite{casamitjana2021synth} and histology-to-CT alignment \cite{leroy2023structuregnet}. Results demonstrate that synthesis can substantially improve cross-modality registration performance. Unpaired image-to-image translation methods, particularly CycleGAN \cite{zhu2017unpaired}, enable cross-modality medical image synthesis without aligned training pairs and have been widely applied to tasks such as MRI-to-CT synthesis \cite{wang2023dc}, stain transfer \cite{vasiljevic2022cyclegan}, and microscopy imaging \cite{tasnadi2023structure}. However, these models are typically trained from scratch and lack strong semantic and structural constraints \cite{wang2024cyclesgan}. Moreover, methods that incorporate structural constraints, such as StructuRegNet \cite{leroy2023structuregnet}, rely on manual contours of rigid anatomical structures, limiting their applicability to soft tissue organs and novel modalities such as HiP-CT. This limits their ability to preserve meaningful cellular morphology across modalities such as H\&E histology and HiP-CT. 

Recent vision foundation models like DINOv2 \cite{oquab2023dinov2} and DINOv3 \cite{simeoni2025dinov3} can learn rich semantic representations through large-scale self-supervised training. This offers promising generalisation to various medical image applications \cite{scholz2025mm,li2025meddinov3}. Although these foundation models generate feature maps of semantic structure and are beneficial to the tasks that require preservation of morphology across domains \cite{baharoon2023evaluating}, deploying a full foundation model independently for each modality is computationally expensive. To address this, Low-Rank Adaptation (LoRA) \cite{hu2022lora} is a lightweight solution to build domain-specific adapters while preserving a single foundation model as a shared encoder.

In this work, our contribution is to propose LoRCA (LoRA Cycle Adaptation), the first DINOv3-based cycle-consistent framework with modality-specific LoRA adapters for H\&E histology to HiP-CT style translation. By combining a shared frozen foundation model backbone with dual LoRA paths, LoRCA enables structure-preserving translation from unpaired data. In this study, we focus on the direction of H\&E histology to HiP-CT, where LoRCA produces volumetrically consistent representations that will facilitate cross-modality registration into the 3D HiP-CT volume. Code is available at https://github.com/UCL-MXI-Bio/2026-zhou-lorca.git.



\section{Method}
\begin{figure}[t]
    \centering
    \includegraphics[width=\linewidth]{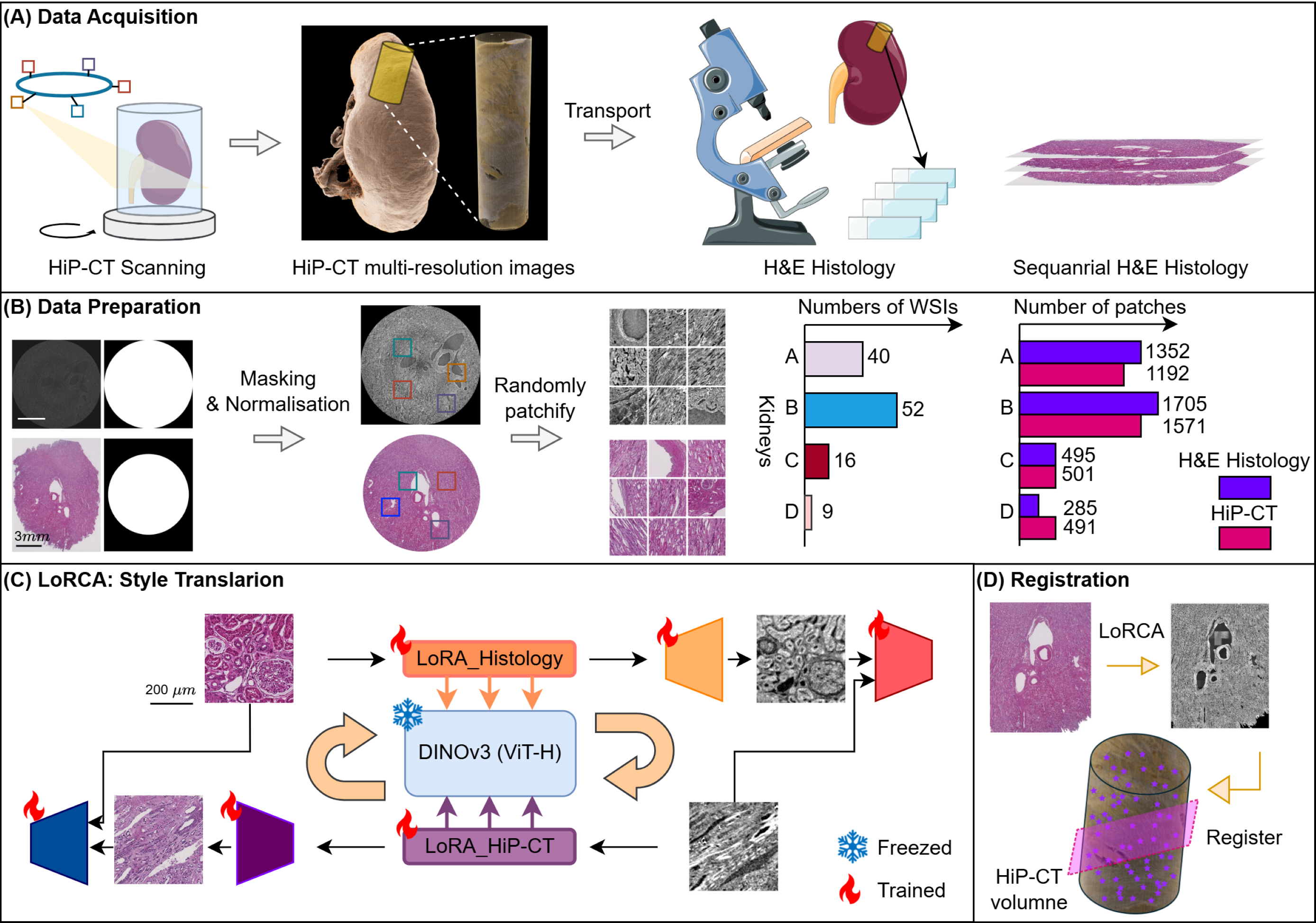}
    \caption{Overview of the pipeline of style translation between H\&E histological sections and HiP-CT. (A) Illustration of organ scanning using HiP-CT, followed by shipping back for histological sectioning on the high-resolution VOIs. (B) Preparations of training data and the statistics of the four training samples. (C) LoRCA structure. (4) Application for registration.}
    \label{fig:blueprint}
\end{figure}
\subsection{Data}
The dataset comprises four human kidneys scanned by HiP-CT, yielding both low-resolution whole-organ volumes (ca. 20 $\mu m$/voxel) and high-resolution volumes of interest (VOIs) (ca. 2 $\mu m$/voxel). The histological sectioning was guided by a 3D-printed mould registered to the high-resolution VOIs. However, precise section-to-volume registration remains challenging due to tissue deformation and reorientation introduced during preparation. Fig. \ref{fig:blueprint} (A) illustrates the data collection procedure. All kidneys were collected following local ethical regulations and approvals: East of England, Cambridgeshire and Hertfordshire Research Ethics Committee (REC 12/EE/0446), and the Laboratoire d’Anatomie des Alpes Françaises. The transportation and imaging protocols received approval from the French Health Ministry.

Serial sectioning was performed on two kidneys, generating 40 and 52 whole slide images (WSIs), respectively, stained with Haematoxylin and Eosin (H\&E), at an x-y resolution of 0.2 $\mu m$ and section thicknesses of 5 $\mu m$. The remaining two kidneys were sectioned non-serially, producing 16 and 9 WSIs, respectively. For each kidney, a HiP-CT subvolume containing all the histological sections was manually identified based on the estimated sectioning depth and anatomical features. Each subvolume consists of 1000-1500 slices, covering 2000 to 3000 $\mu m$ along the z-axis. To match the histological coverage, we uniformly sample HiP-CT slices at twice the number of histological sections from each specimen. 

As shown in Fig. \ref{fig:blueprint} (B), before dataset construction, we applied image normalisation, resampled to 2 $\mu m$, and generated 256$^2$ pixel patches randomly, with patches containing less than 90\% tissue content discarded. The training and test patch datasets were split based on an 80:20 ratio at the sample level.

\subsection{Shared DINOv3 Backbone with Dual LoRA Paths}
LoRCA is a cycle-consistent 2D generative framework for unpaired cross-modality translation between RGB H\&E histology and grayscale HiP-CT. As shown in Fig. \ref{fig:blueprint} (C), rather than relying on task-specific encoders trained from scratch, LoRCA leverages a shared frozen DINOv3 backbone, which provides rich semantic and structural representations that are difficult to obtain from limited biomedical training data. To adapt the shared backbone across two modalities without full fine-tuning, modality-specific dual LoRA adapters are injected into all linear layers of the Vision Transformer (ViT). The extracted features are decoded back to image dimensions through separate UNet-style decoders, then fed to PatchGAN discriminators. Cycle consistency is employed as the training loss to enforce structure-preserving translation without paired supervision.

Formally, for a frozen linear layer with an input $\mathbf{x}$ and a weight matrix $\mathbf{W} \in \mathbb{R}^{d \times k}$, the LoRA-adapted output $\mathbf{h}$ is computed as:

\begin{equation}
\mathbf{h} = \mathbf{W}\mathbf{x} + \frac{\alpha}{r}\mathbf{B}\mathbf{A}\mathbf{x} + b,
\end{equation}
where $\mathbf{A} \in \mathbb{R}^{r \times k}$ and $\mathbf{B} \in \mathbb{R}^{d \times r}$ are trainable low-rank matrices with adapter rank $r$, a scaling factor $\alpha$ and a bias term $b$. The backbone weights $\mathbf{W}$ remain frozen. In LoRCA, two independent sets of LoRA parameters $\{\mathbf{A}^{hipct}, \mathbf{B}^{hipct}\}$ for HiP-CT and $\{\mathbf{A}^{hist}, \mathbf{B}^{hist}\}$ for H\&E are maintained, while all other backbone parameters are shared and frozen. During training, the LoRA adapter for the input modality is activated, producing modality-specific feature representations $\mathbf{F}^{hipct}$ and $\mathbf{F}^{hist}$ from the same DINOv3 backbone.

\subsection{UNet-Style Decoders}
Each modality has a dedicated UNet-style decoder, as shown in Fig. \ref{fig:blueprint} (C), that upsamples the DINOv3 patch token representations back to image dimensions. The decoding process includes a series of bilinear upsamplings and convolutional layers. The decoder for HiP-CT generation produces single-channel grayscale output, while the decoder for generative H\&E histology produces three-channel RGB output, reflecting the respective modality characteristics. We evaluated different connection types: plain, fusion, and skip connections between DINOv3 and the decoder, and report them in Sect. \ref{results}.

\subsection{Cycle-consistent Training}
LoRCA adopts a cycle-consistent adversarial training strategy to style translation in the absence of paired data. Two translation paths are defined: histology to HiP-CT ($G_{hist \rightarrow hipct}$) and HiP-CT to histology ($G_{hipct \rightarrow hist}$), each consisting of the shared DINOv3 backbone with its corresponding LoRA adapter and modality-specific decoder. Two PatchGAN discriminators $D_{hipct}$ and $D_{hist}$ provide adversarial supervision to encourage realism in the translated outputs.

The total training objective combines adversarial and cycle consistency losses:
\begin{equation}
\mathcal{L}_{total} = \lambda_{adv}(\mathcal{L}_{adv}(G_{hist \rightarrow hipct}, D_{hipct}) + \mathcal{L}_{adv}(G_{hipct \rightarrow hist}, D_{hist})) + \lambda_{cyc} \mathcal{L}_{cyc},
\end{equation}
where the adversarial loss $\mathcal{L}_{adv}$ follows the least-squares GAN formulation \cite{mao2017least}. $\lambda_{adv}$ and $\lambda_{cyc}$ are the relative weights of the adversarial and cycle consistency terms. The cycle consistency loss is:
\begin{equation}
\begin{split}
\mathcal{L}_{cyc} &= \mathbb{E}\left[|G_{hipct \rightarrow hist}(G_{hist \rightarrow hipct}(\mathbf{x}^{hist})) - \mathbf{x}^{hist}|_1\right] \\ 
&+ \mathbb{E}\left[|G_{hist \rightarrow hipct}(G_{hipct \rightarrow hist}(\mathbf{x}^{hipct})) - \mathbf{x}^{hipct}|_1\right].
\end{split}
\end{equation}

\section{Results}
\label{results}
In this work, LoRCA applies LoRA adapters with a rank $r=8$ and a scaling factor $\alpha=16$. For training loss, $\lambda_{adv}$ and $\lambda_{cyc}$ are set to 5 and 10 respectively. All the learning rates for LoRA adapters and discriminators are set to 0.0003, and training is performed on a single NVIDIA A100 with a batch size of 8.

\subsection{Generator Structure Selection}

\begin{figure}[t]
    \centering
    \includegraphics[width=\linewidth]{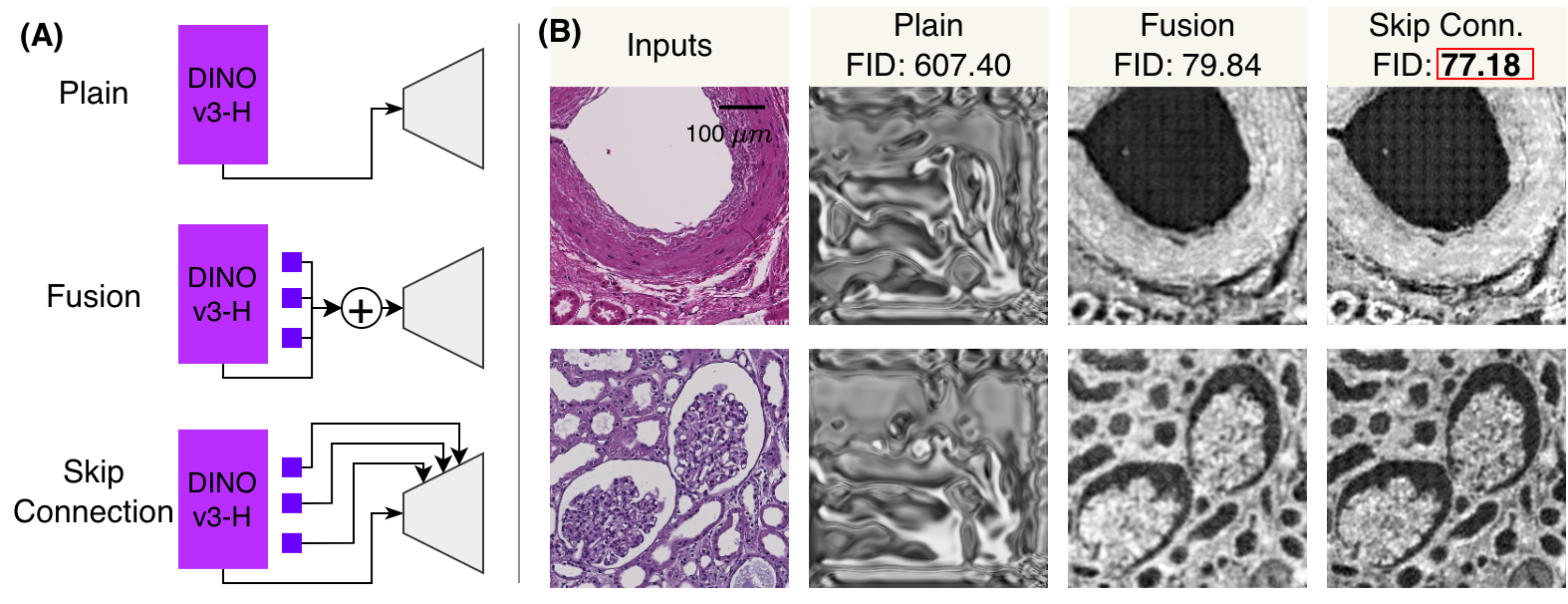}
    \caption{LoRCA generator structure selection. (A) Illustration of different generator structures. (B) Generative HiP-CT from different models on H\&E histology validation patches: vasculature wall (first row) and glomeruli (second row).}
    \label{fig:structure_selection}
\end{figure}

Firstly, we conducted an ablation study to determine the optimal generator architecture. Three generator structures were evaluated: a plain decoder that uses only the final patch tokens, a fusion decoder that aggregates intermediate features via summation, and a skip-connection decoder that concatenates intermediate features to the corresponding decoder stages. Unlike convolutional backbones, DINOv3 is built on a ViT that produces patch tokens at a fixed resolution rather than hierarchical feature maps. For skip-connection structures, intermediate outputs are upsampled to construct feature representations. To evaluate the generated effects, FID was computed between real HiP-CT training patches and generated HiP-CT patches translated from histology. As shown in Fig.~\ref{fig:structure_selection}, the plain decoder produces heavily distorted outputs (FID: 607.40), indicating that final patch tokens alone result in a model failure mode and are insufficient to recover spatial detail without hierarchical features. The fusion decoder substantially improves realism (FID: 79.84), while the skip-connection decoder achieves the best FID of 77.18 with the most structurally faithful outputs across all tissue regions. Based on these results, the skip connection architecture is adopted for all subsequent experiments.

\subsection{Histology-to-HiP-CT Translation Quality}

\begin{table}[t]
\caption{Evaluation of the generated HiP-CT quality from histology on the training and validation datasets}
\label{tab:eval_gen_hipct}
\centering
\begin{tabular*}{\linewidth}{@{\extracolsep{\fill}}lcccc}
\toprule
 & \multicolumn{2}{c}{Mutual Information} & \multicolumn{2}{c}{Canny Edge IoU} \\
\cmidrule(lr){2-3} \cmidrule(lr){4-5}
Methods & Train & Test & Train & Test \\
\midrule
CycleGAN (ResNet-9) & 0.744 $\pm$ 0.077 & 0.741 $\pm$ 0.073 & 0.188 $\pm$ 0.024 & 0.188 $\pm$ 0.024 \\
CycleGAN (UNet256)  & 0.743 $\pm$ 0.054 & 0.743 $\pm$ 0.050 & 0.167 $\pm$ 0.032 & 0.166 $\pm$ 0.030 \\
LoRCA (Skip conn.)  & \textbf{0.834 $\pm$ 0.088} & \textbf{0.831 $\pm$ 0.084} & \textbf{0.418 $\pm$ 0.052} & \textbf{0.417 $\pm$ 0.051} \\
\bottomrule
\end{tabular*}
\end{table}

\begin{figure}[t]
    \centering
    \includegraphics[width=\linewidth]{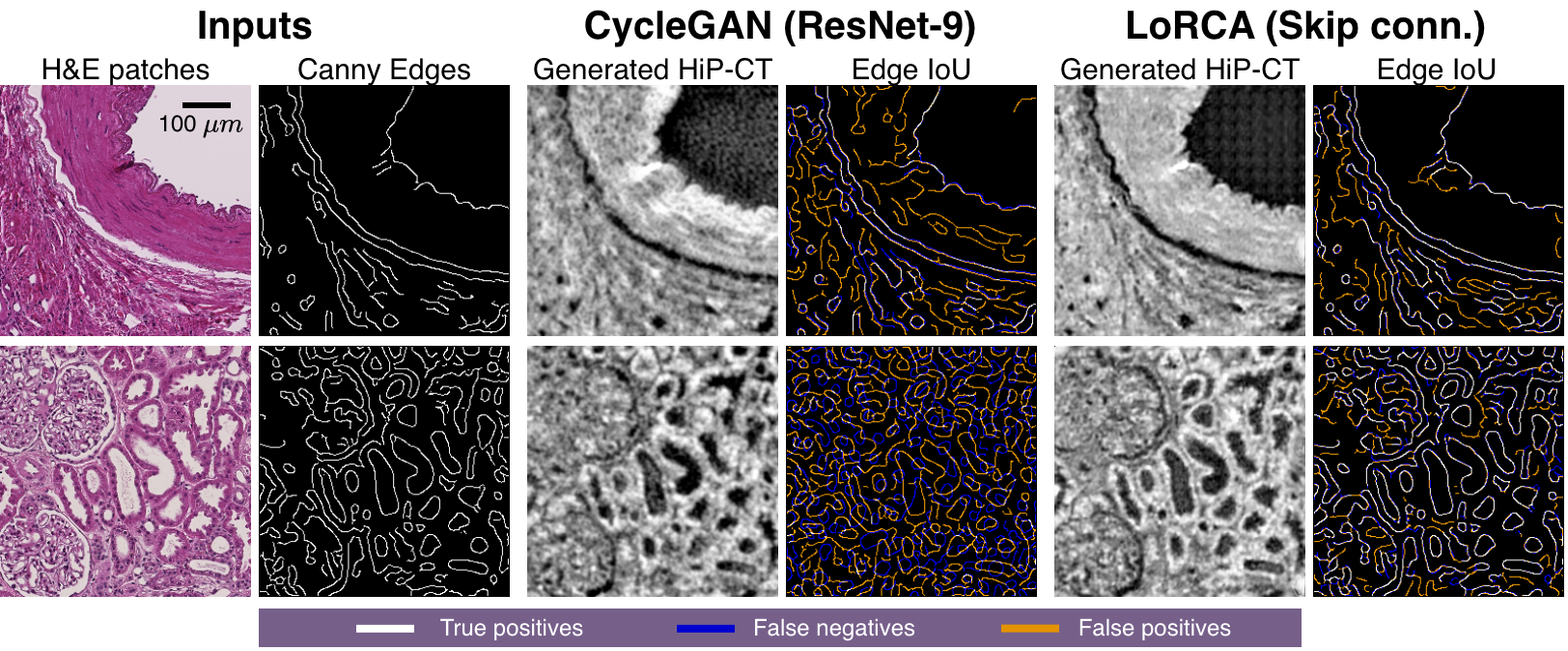}
    \caption{Comparison of edge preservation for CycleGAN and LoRCA on representative patches containing a blood vessel (top row) and a cortex with glomeruli (bottom row). Input Canny edge maps and edge overlap visualisations from generative ones are shown.}
    \label{fig:edge_comparison}
\end{figure}

Table~\ref{tab:eval_gen_hipct} shows the mutual information (MI) and intersection over union (IoU) of the Canny edge detection results on the generated HiP-CT patches and the corresponding H\&E histology patches on both training and test datasets. The results indicate that LoRCA (skip conn.) consistently outperforms CycleGAN baselines across metrics and splits, achieving higher MI scores compared with CycleGANs. The Canny edge IoU also improves substantially from 0.188 (ResNet-9) and 0.166 (UNet256) to 0.417 for LoRCA on test dataset, representing a more than twofold improvement in structural edge preservation. Similar performance on training and test datasets suggests good generalisation without overfitting.

Qualitative results are shown in Fig.~\ref{fig:edge_comparison}, where two representative H\&E patches, which cover distinct tissue regions: a vessel wall and a renal cortex containing glomeruli, are style-translated by each method. The Canny edge overlay compares structural boundaries between the generated HiP-CT and the input H\&E patch, with white indicating edges that overlap between the original input and the style-translated image (true positives), blue indicating edges present in the input but missing in the generated output (false negatives), and orange indicating hallucinated edges in the generated output absent from the input (false positives). CycleGAN (ResNet-9) produces outputs with reasonable global contrast but fails to faithfully recover the fine structural boundaries present in the H\&E input. Its edge maps show extensive blue regions reflecting missed vessel walls and glomerular capsules, alongside orange regions from hallucinated structures. In contrast, LoRCA produces substantially more edge overlays (in white) across both tissue regions, indicating that the DINOv3-derived features enable more faithful transfer of structural boundaries from the H\&E input into the generated HiP-CT appearance. The vessel wall example (top row) demonstrates LoRCA's ability to preserve large-scale tissue interfaces, while the glomerular (bottom row) example highlights its advantage at the cellular scale, where precise boundary preservation is most critical for downstream slice-to-volume registration.

\subsection{Feature matching}
\begin{figure}[t]
    \centering
    \includegraphics[width=\linewidth]{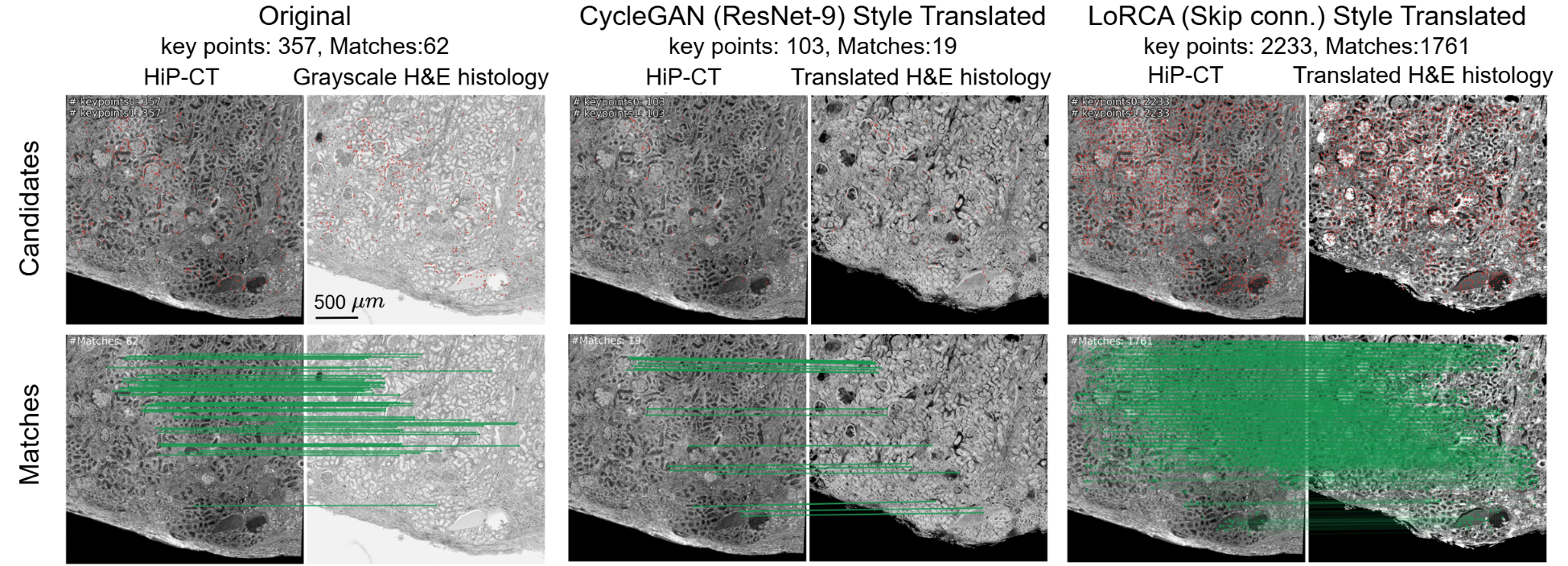}
    \caption{Feature points searching and matching on manually aligned HiP-CT slice and histological WSI using MatchAnything \cite{he2025matchanything}.}
    \label{fig:registration}
\end{figure}

To provide an initial assessment of whether the improved image translation quality and edge preservation of LoRCA translate into better cross-modality registration, keypoint matching was evaluated between a test H\&E slice, represented either as a grayscale image, the CycleGAN-translated image, or the LoRCA-translated image, and its corresponding HiP-CT slice, which had been accurately aligned manually, as described in \cite{walsh2021imaging}. Correspondences were computed using MatchAnything \cite{he2025matchanything}, a recently developed feature matching framework that identified equivalent image structures across modalities. For each image, MatchAnything first identified candidate keypoint correspondences, after which Random Sample Consensus (RANSAC) removed geometrically inconsistent matches. 

The results shown in Fig.~\ref{fig:registration} demonstrate a substantial improvement in cross-modality feature matching with LoRCA. Direct matching between grayscale histology and HiP-CT yielded 357 candidate correspondences, of which only 62 were retained after RANSAC (17.4\%). CycleGAN performed worse, yielding 103 candidate correspondences and 19 retained matches (18.4\%). In contrast, LoRCA produced 2,233 candidates, with 1,761 retained (78.9\%), representing a more than 28-fold increase in reliable matches relative to the grayscale histology.

\section{Conclusion and limitations}
This work presented LoRCA, a DINOv3-based cycle-consistent framework with modality-specific LoRA adapters for histology to HiP-CT translation. Compared with CycleGAN, LoRCA achieved higher structural fidelity and enhanced feature matching. However, limitations include the strategies for intensity normalisation of both modalities to overcome the over-exposure effects on generative images and the conversion of grayscale HiP-CT to three-channel inputs for DINOv3. The current evaluations only focus on style translation quality and feature correspondence without registration accuracy, and we have not yet isolated backbone freezing's contribution via a controlled ablation. Future work will integrate LoRCA into a full slice-to-volume registration framework such as MatchAnything \cite{he2025matchanything} or StructuRegNet \cite{leroy2023structuregnet} and quantify its impact on registration accuracy.

\begin{credits}
\subsubsection{\ackname}
This work was supported by the Chan Zuckerberg Initiative DAF, an advised fund of Silicon Valley Community Foundation (DAF2022-316777 and CZIF2024-009938), NIH BRAIN Initiative CONNECTS program via National Institute of Neurological Disorders and Stroke (NINDS) and National Institute of Mental Health (NIMH) (UM1-NS132358), Wellcome Trust (310796/Z/24/Z), Royal Academy of Engineering (RAEng) (CiET1819/10), and CIFAR MacMillan Multiscale Human Fellowship. We acknowledge the ESRF and the HOAHub for provision of synchrotron radiation facilities under UCL-led proposals MD-1290, MD-1389 and LS-3686 on beamlines BM18. We also sincerely thank David Stansby (Data Scientist, UCL) and Guillaume Gaisné for their support with data management, as well as Joanna Purzycka and Jean-Pascal Hograindleur for sample preparation and mounting. The authors also sincerely thank those who donated their bodies to science, enabling anatomical research that can potentially increase humanity’s overall knowledge and then improve patient care. Therefore, these donors and their families deserve our highest gratitude. 

\end{credits}

%
%
%
\bibliographystyle{splncs04}
\bibliography{references}
\end{document}